\documentclass[preprint,prl]{revtex4}
\usepackage{graphicx}
\usepackage{textcomp}
\usepackage{amsmath}
\usepackage{fontenc}
\usepackage{subfigure}
\usepackage{tabularx}  
\usepackage{float}
\usepackage[english]{babel}
\usepackage[active]{srcltx}

\begin{document}

\title{Unwinding of the twist-bend nematic structure under an external field}
\author{R. S. Zola}\email{Corresponding Author: rzola1@kent.edu}
\affiliation{Universidade Tecnol\'ogica Federal do Paran\'a, Campus Apucarana,\\
Rua Marc\'ilio Dias 635, 86812-460 Apucarana, Paran\'a, Brazil \\}

\author{G. Barbero}
\affiliation{Department of Applied Science and Technology,
Politecnico di Torino, Corso Duca degli Abruzzi 24, 10129 Torino, Italy.}

\author{I. Lelidis}
\affiliation{Solid State Section, Department  of Physics, University of Athens,\\
Panepistimiopolis, Zografos, Athens 157 84, Greece.}

\author{M. P. Rosseto}, \author{L. R. Evangelista} 
\affiliation{Departamento de F\'isica, Universidade Estadual de Maring\'a\\ Avenida Colombo, 5790-87020-900 Maring\'a, Paran\'a, Brazil}

\begin{abstract}
 The recently discovered twist-bend nematic phase, $\rm N_{tb}$, is a non-uniform equilibrium nematic phase that presents a spontaneous bend with a precession of the nematic director, \textbf{n}, on a conical helix with a tilt angle $\theta$ and helical pitch $P$. The stability of the $\rm N_{tb}$  phase has been recently demonstrated from the elastic point of view by extending the Frank elastic energy density of the nematic phase to include the symmetry element of helical axis, $ \textbf t$. In the present communication, we investigate the influence of an external bulk field (magnetic or electric) on the pitch of the $\rm N_{tb}$ phase. For a material with positive magnetic susceptibility anisotropy, when the magnetic field $\textbf{H}$ is parallel to the helical axis, $\textbf{t}$, the field modifies the coupling parameter between the director and the helical axis, thus shifting the interval of values for which this coupling results in the $\rm N_{tb}$ phase. If $\textbf{H}$ is perpendicular to $\textbf{t}$, it is shown that the pitch of the nematic director $\textbf n$ precession increases with $\textbf{H}$, and at a critical value of $H_c$ it becomes infinite as in a cholesteric liquid crystal, that is, one gets a uniform orientation of n.  Our results include the behaviour of a cholesteric under field as a limiting case. 
\end{abstract}
\maketitle

\section{I. Introduction}

Liquid crystals (LCs) are a fascinating example of collective behavior in nature. In contrast to ordinary liquids, the orienting effect over the anisometric molecules by an external influence (such as a magnetic field)  is proportional to the volume of the responding unit of LC molecules~\cite{Coll}, resulting in strong aligning effect for relatively small fields. Whenever an external field is used to align or reorient a LC media, a distortion in the orientation may occur~\cite{freed}. Such astonishing effect is the backbone of modern electro-optic devices (displays), where the average orientation of the molecules results from the competition among external field, anchoring strength and elasticity~\cite{deng}. When the system is chiral, an applied field yields a richer class of textures and transitions when compared to the non-chiral counterpart, such as the appearance of fingers~\cite{f1,f2,f3,f4}, bubble domains~\cite{b1,b2}, and many others~\cite{b3,a2}. One important example is the field induced (magnetic, electric, mechanical) cholesteric to nematic transition, where the cholesteric helix is unwind when the field is applied perpendicularly to the helix direction for positive dielectric anisotropy materials both for an unbound sample~\cite{degennes1,meyer2} and in confined geometry~\cite{scarfone,lelidis}. This unwinding is specially interesting in the vicinities of the transition cholesteric to Smectic A phase~\cite{Jamee}. 

Recently, a new class of materials has been discovered in the LC field with potential for presenting exciting orientational transitions as the ones aforementioned. The twist bend nematic ($\rm N_{tb}$) phase is a heliconical structure with a nanoscale pitch favored by bent-shaped molecules. It has been theoretically predicted years ago~\cite{Meyer, Dozov, Memmer}, but only recently the $\rm N_{tb}$ phase  was experimentally observed  in bent molecular dimers~\cite{Oleg,Adlem, Noel}, trimers~\cite{quan} and, recently, in  rigid bent-core materials~\cite{Min} and in chiral dimers as well~\cite{chiraldop1}. Sparked by these experimental achievements, new theoretical models have been published  recently \cite{sm1,sm2,virga}, contributing to better understanding $\rm N_{tb}$ materials. In this phase, the ground state presents a twist-bend deformation with the director precessing around a helix axis maintaining a constant angle $0\leq\theta\leq\pi/2$~\cite{Oleg}. From the optical scale point of view, the $\rm N_{tb}$ phase forms micron size stripes resembling smectics. However, x-ray measurements have shown no periodic variation in the electron density~\cite{Adlem}. In fact, these stripes have been attributed to a Helfrich-Hurault-type deformation caused by the undulation of the pseudolayers forming the phase~\cite{Helfrich,Hurault,Challa}. 

\begin{figure}[htp]
\centering
\includegraphics[scale=0.3,clip]{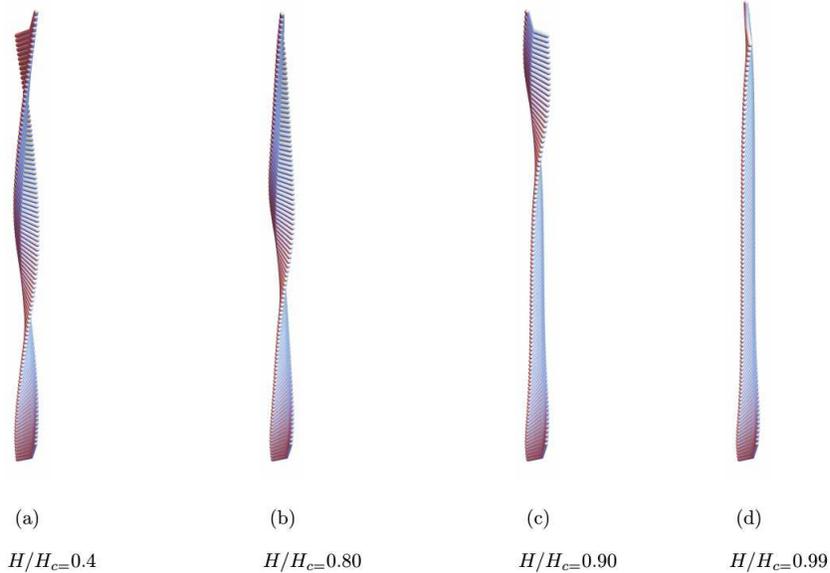}
\caption{Calculated director profile for several ratio $H/H_c$, made by discretizing the resulting equations in space. a)$H/H_c=0.40$, b)$H/H_c=0.80$, c)$H/H_c=0.90$ and d)$H/H_c=0.99$. }
\label{fig01}
\end{figure}

In a recent work, Challa et. al.~\cite{Challa} studied the effect of high magnetic fields on an $\rm N_{tb}$ material. Nonetheless, their focus was centered on the temperature shift of the $\rm N_{tb}$ to nematic caused by the field. They observed however that the optical stripes disappeared under the field. Similar experiments have been conducted in~\cite{Oleg}, but this time under electric field, where a first order like transition from the homeotropic to $\rm N_{tb}$ texture happens. However, there is no data so far on the unwinding of the $\rm N_{tb}$ helix. In this article, we   
take advantage of the recent elastic model for the phase~\cite{pre} and calculate the pitch's dependence with an applied magnetic field. If the field is applied parallel to helical axis, it modifies the coupling parameter between the director and the helical axis, thus shifting the interval of values for which this coupling results in the $\rm N_{tb}$ phase. If applied perpendicular, it is shown that the pitch increases with $H$ (magnetic field). At a critical value $H_c$, the pitch is unwounded and an uniform director distribution 
occurs~\cite{degennes1,meyer2}. In figure~\ref{fig01}, the calculated director profile for different $H/Hc$ ($H_c$ is the critical field) is presented in order to illustrate the phenomenon, made by discretizing the resulting equations. As the field is increased, the helix unwinds until an uniform nematic phase forms. In these curves, each small bent-rod represents the director orientation at a given position in space. The bend in the rods is made intentionally in order to show the symmetry of the material. Our results  include the behavior of a cholesteric under an applied field as a limiting case.


\section{The elastic energy density }

One of the most remarkable points regarding the $\rm N_{tb}$ phase is the fact that it presents, in the ground state, spontaneous twist and bend deformations~\cite{Noel}. Recently, we have proposed a new elastic theory, in a very general form by using the symmetry point of view, that accounts for the $\rm N_{tb}$ phase~\cite{pre}. It was shown the stability of the phase and that $K_{22}>K_{33}$ is a requirement, although there is no need to assume negative elastic constants~\cite{pre}. 
In the $\rm N_{tb}$ phase the nematic director $\mathbf{n}$ forms a constant tilt angle say $\theta$ in respect to the helical axis direction $\mathbf{t}$ and it turns around $\mathbf{t}$.
Taking ${\bf t} = {\bf u_z}$, the nematic director is written as

\begin{equation}
\centering
{\bf n}= [\cos \phi (z) \,{\bf u_x}+ \sin \phi (z) \,{\bf u_y }] \sin \theta + \cos \theta\, {\bf u_z}.
 \label{1a}
\nonumber
\end{equation}
As shown in reference~\cite{pre}, this director configuration simplifies the free energy in such a way that only the following terms will contribute to energy density, namely:

\begin{eqnarray}
 &f_d= f_1 - \frac{1}{2}\eta ({\bf n} \cdot {\bf t})^2 + \frac{1}{2}K_{22}[{\bf n} \cdot (\nabla \times {\bf n)}+q_0]^2\nonumber \\
 &+ \frac{1}{2} K_{33} ({\bf n} \times \nabla \times {\bf n})^2+ \nu_4[({\bf t} \cdot \nabla){\bf n}]^2,
 \label{a1}
\end{eqnarray}
where $f_1=f_0-(1/2)\, K_{22}q_0^2$ and $q_0=\kappa_2/K_{22}$. Notice that $f_1$ is a constant. From 
equation~\ref{a1}, it is clear that $q_0$ represents a natural twist of the system, as it is well known for regular cholesterics. In order to investigate the effect of an external magnetic (electric) field on an unbounded $\rm N_{tb}$  liquid crystal, we consider the elastic part of the free energy, equation~\ref{a1}, and the magnetic energy term describing the coupling between the nematic director and the external magnetic field, $\bf{H}$.
Substituting this expression for ${\bf n}$ (equation~\ref{1a}) in the expression for the free energy, one has

\begin{equation}
 f_d(\phi', x)=f_{el}+f_{mag},
 \label{5}
\end{equation}
in which the elastic part reduces to the simple form

\begin{eqnarray}
f_{el}(\phi', x) = f_1 &-& \frac{1}{2}\eta (1 - x) + \frac{1}{2}K_{22} (qx-q_0)^2 \nonumber \\
&-& \frac{1}{2} K_{33}q^2(x^2-xb_0)
\label{6a}
\end{eqnarray}
and
\begin{equation}
 f_{mag}=-\frac{1}{2}\chi_a[{\bf H} \cdot {\bf n}]^2,
 \label{6b}
\end{equation}
with $x=\sin^2 \theta$, $\phi=qz$, $b_0=(1+2\nu_4/K_{33})$, and $\chi_a$ is the magnetic susceptibility anisotropy. In a perfectly aligned nematic phase, $x=0$ and therefore no bend distortion occurs. In the cholesteric phase, $x=1$, and the medium will have no extra distortion only if $b_0=1$. In fact, by setting $\partial f_d/\partial q=0$, in the absence of the external field we find the equilibrium wave number, given by

\begin{eqnarray}
 q^*=\pm\frac{K_{22}q_0}{b_0K_{33}+K_{22}x-K_{33}x}.
 \label{5}
\end{eqnarray}
From equation~\ref{5}, we notice that when $x=1$ (cholesteric phase), $q^*=q_0$ if $b_0=1$ as mentioned above. Thus, $q_0$ refers to the regular cholesteric wave number. When $x=0$ (nematic phase), $q_{tp}^*=K_{22}q_0/K_{33}$ (again, $b_0=1$) which corresponds to the transient planar wave number and represents the pitch right after the transition from the homogenous (nematic) texture to the cholesteric one~\cite{transient}. 
Notice however that recent measurements indicate that $K_{22}$ is at least one order of magnitude larger than $K_{33}$ \cite{Adlem} and, therefore, the transient pitch for these materials should be smaller than the natural pitch. In the $\rm {N_{tb}}$ phase, $\mid q^*\mid>\mid q_0\mid$ for $b_0\neq 1$ and $0< x< 1$.  This sets the range of validity of $b_0$, or $b_0<K_{22}/K_{33}+x(1-K_{22}/K_{33})$. This explicit dependence that the pitch length has on $b_0$ implies that the term related to $b_0$ in the free energy, eq.~\ref{a1}, must not be neglected in the $\rm {N_{tb}}$ phase. In fact, this could be explained by a different arrangement of the medium when compared to regular cholesterics when $b_0\neq 1$. 
By setting $\partial f_d/\partial x=0$, we find

\begin{eqnarray}
 x^* &=&-\frac{b_0K_{33}\mp K_{22}q_0\sqrt{b_0K_{33}/\eta}}{K_{22}-K_{33}}\quad {\rm and} \nonumber \\
 q^* &=&\pm\frac{\sqrt{\eta}}{\sqrt{b_0K_{33}}}.
 \label{stars}
\end{eqnarray}
 Both $q^*$ and $x^*$ correspond to the equilibrium configuration only if $\partial^2f/\partial x^2\geq0$, or $K_{22}-K_{33}\geq 0$. In fact, by replacing $q^*$ in $x^*$ in equation~\ref{stars}, one obtains
 
\begin{equation}
 x^*=\frac{K_{22}q_0/q^*-b_0K_{33}}{K_{22}-K_{33}}.
 \label{7n}
\end{equation}
From equation $\ref{7n}$, when $q^*=q_0$, $x^*=1$ (cholesteric) only when $b_0=1$. If $b_0\ne1$, then 
$x^* = 1 - 2\nu_4/(K_{22}-K_{33})$. Therefore, $b_0\neq 1$ represents an extra deformation, which in the cholesteric phase causes a tilt with respect to the helical axis. Furthermore, in this cholesteric phase, $b_0$ must be always larger than 1. 
 
Hereafter, the total energy of the system will be written as

\begin{equation}
\centering
F=\int_V f_d d^3 r,
\end{equation}
and we will analyze two possible situations separately. Firstly, we consider the case in which the field is applied parallel to helical axis and, after that, the case in which the field is applied perpendicularly to the helix. 

\section{First case: field parallel to helix axis}

\begin{figure}[htp]
\centering
\includegraphics[scale=0.3,clip]{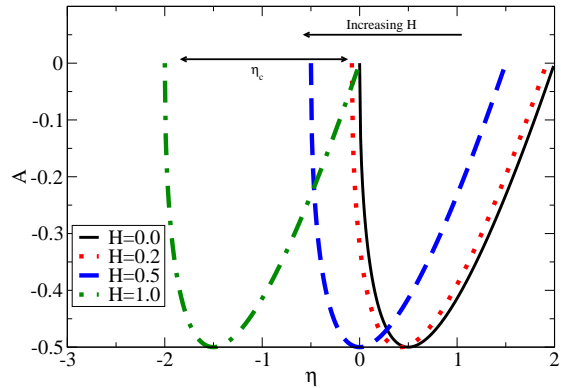}
\caption{Behavior of $A$ versus $\eta$ for different values of strength of the field ${\bf H}$.  For simplicity, we choose $K_{22}=K_{33}=q_0=1$, $\chi_a=2$ and $b_0=0.5$. }
\label{figure:Fig2}
\end{figure}

If ${\bf H}\parallel{\bf n}$, a realigning torque is applied to the director. In this situation, one expects a complicated reorientation process to occur, which will eventually lead to a nematic, homogeneous 
state~\cite{Oleg}. For typical cholesteric samples, the field for a bounded sample drives the texture, originally planar anchored, to the so-called focal conic texture (or the Helfrich deformation), followed by the finger print texture until the helix is completely unwind~\cite{deng}. This situation demands the angle $\theta$ to change across the sample as a function of the field. However, in our case, due to the complexness of the calculations, we will study the case for a uniform $\theta$ along the sample and analyze the role of the field on shifting the transition from $\rm N_{tb}$ to $N$~\cite{Challa}. To start, we notice that in this situation equation~\ref{6b} becomes

\begin{equation}
f_{mag}=-\frac{1}{2}\chi_aH^2(1-x).
\end{equation}
Subsequently, one notices that,  under an applied field, the equilibrium state found by imposing $\partial f_d/\partial x=0$, $\partial f_d/\partial\phi = 0$ and $\phi=qz$ ($q$ is the wavenumber) is given by~\cite{pre}:

\begin{eqnarray}
x^*(\eta) &=&-\frac{b_0K_{33}\mp K_{22}q_0\sqrt{\frac{b_0K_{33}}{\eta+\chi_aH^2}}}{K_{22}-K_{33}}\quad {\rm and} \nonumber \\
q^*(\eta) &=&\pm\frac{\sqrt{\eta+\chi_aH^2}}{\sqrt{b_0K_{33}}}.
 \label{7}
\end{eqnarray}
Therefore, both $q^*(\eta)$ and $x^*(\eta)$, representing the equilibrium state, are a function of the applied field. Now, we analyze the results by rewriting equation~\ref{5} in the Landau-like form

\begin{equation}
 f(x)=f_0-\frac{\eta\chi_aH^2}{2}+Ax+Bx^2, 
\end{equation}
where $ A = -K_{22}q_0q^*(\eta)+ \eta/2+ \chi_aH^2/2+K_{33}b_0q^{*2}(\eta)/2 $ and $B = (K_{22}-K_{33}) q^*(\eta)^2/2$. The twist-bend phase, which corresponds to $x\neq 0$, is favored from the energy point of view
only if $A<0$. Thus, there is a second order phase transition from the usual nematic phase ($x=0$) to a twist-bend nematic ($x\neq 0$) when $A=0$. The parameter $\eta$, whose value quantifies the coupling between the director and the helical axis and, thereby, the stability of the $\rm N_{tb}$ phase itself, has a critical value at this point. Therefore, by setting $A=0$, the interval of possible values for $\eta$ that results in a stable  $\rm N_{tb}$ phase is found to be

\begin{equation}
 \eta_{c1}=-\chi_aH^2 \quad {\rm and} \quad 
 \eta_{c2}=\eta_{c0}-\chi_aH^2,
 \label{8}
\end{equation}
where $\eta_{c0}=4K_{22}^2q_0^2/b_0K_{33}$ is the critical value in the absence of external 
field~\cite{pre}. From equation~\ref{8}, it is clear that the effect of the applied field is to shift to lower values the interval of stability of the $\rm N_{tb}$ phase. It is worth noticing that the range interval is always the same, $\eta_c$, but as the field is increased this interval is shifted. This could be used to explain, at least qualitatively, the results found in~\cite{Challa}, where the applied magnetic field lowers the transition temperature from the nematic to the twist-bend phase, understood from the point of view of our model as caused by shifting the range of stability of the phenomenological parameter $\eta$.
Figure~\ref{figure:Fig2} shows the parameter $A$ as a function of $\eta$. As the strength of the field ${\bf H}$ is increased, the curves shift to the left. 

\section{Second case: field perpendicular to helix axis}
In the following,  we suppose that the magnetic field $\bf{H}$ (i) is applied along the \textit{y}-direction, that means it is perpendicular to the helical axis {\bf t}, and (ii) it does not change the tilt angle $\theta$. In this case, equation~(\ref{5}) is casted in the form:

\begin{eqnarray}
\centering
f(\phi', x) &=& f_1 - \frac{1}{2}\eta (1 - x) + \frac{1}{2}K_{22} (\phi'x-q_0)^2 \nonumber \\
&-& \frac{1}{2} K_{33}\phi'^2(x^2-xb_0)-\frac{1}{2}\chi_a H^2x^2 \sin^2\phi. \nonumber
\end{eqnarray}

In order to calculate $\phi(z)$ that minimizes the energy and taking into account that there is a 1D periodicity in the system, one has to minimize the energy per period~\cite{Priestley}, that means, the functional

\begin{equation}\label{FA}
\centering
\frac{F(\lambda)}{S}= \int_0^\lambda dz \left[\frac{\gamma}{2} - K_{22}q_0 x \phi'+ \frac{1}{2}\alpha \phi'^2 -\frac{1}{2}\chi_a H^2x^2 \sin^2\phi \right],
\nonumber
\end{equation}
where $S$ is the area of the system in the $(x,y)$ plane, $\gamma=2 f_1 - \eta (1 - x)+K_{22}q_0^2$, and 
$\alpha=-K_{33}(x^2-xb_0)+K_{22}x^2$ is an effective elastic constant. The corresponding Euler-Lagrange equation is

\begin{equation}
\centering
\xi^2 \frac{d^2 \phi}{dz^2}+ \cos \phi \sin\phi=0,
\label{EL}
\end{equation}
with $\xi= \frac{\sqrt{\alpha/\chi_a}}{H x}$ the magnetic coherence length~\cite{jacgil}. The Euler-Lagrange 
equation~(\ref{EL}) has the first integral

\begin{equation}
\centering
\left(\frac{d\phi}{dz}\right)^2=\frac{1}{\xi^2}\left(\frac{1}{C ^2}-\sin^2\phi\right),
\end{equation}
where $C$ is an integration constant. Finally,

\begin{equation}
\centering
\frac{d\phi}{dz}=\pm\frac{1}{C\xi}\sqrt{1-C^2\sin^2\phi}, 
\nonumber
\end{equation}
where the signs refer respectively to the positive and negative helicity. The wavelength of the modulations is then given by

\begin{equation}\label{lambda}
\centering
\lambda=\int^\lambda _0 dz = 2 \xi CE_1(C),
\end{equation}
with 

\begin{equation}
\centering
E_1(C)=\int^\frac{\pi}{2}_0 \frac{d\phi}{\sqrt{1-C^2\sin^2\phi}}
\nonumber
\end{equation} 
being the elliptic integral of the first kind.

For further calculation, we
introduce the average energy density per period, $g$,  defined as

\begin{equation}
\centering
g=\frac{2F(\lambda)}{S\lambda},
\nonumber
\end{equation}
that, taking into account equation~(\ref{FA}) and after some algebra, assumes the following form:

\begin{equation}
\centering
g= \gamma- \frac{2K_{22}q_0 x \pi}{\lambda}- \frac{\alpha}{C^2 \xi^2} +\frac{2\alpha}{\lambda C \xi} \int^\pi_0 \sqrt{1-C^2 \sin^2\phi}d\phi.
\label{g2}
\end{equation}

Minimization of $g$ yields
\begin{equation}
\centering
\frac{dg}{dC^2}= \frac{d\lambda}{d C^2}\left[\frac{2\pi K_{22}q_0 x}{\lambda^2}-\frac{2\alpha}{\xi C \lambda^2}\int^\pi_0 \sqrt{1-C^2\sin^2\phi}d\phi \right]=0
\nonumber
\end{equation}
which allow us to calculate $C(H)$

\begin{equation}
\centering
\dfrac{C}{E_2(C)}=\frac{2\sqrt{\alpha\,\chi_a}}{\pi K_{22} q_0}\,H,
\label{1}
\end{equation}
where

\begin{equation}
\centering
E_2(C)=\int^\frac{\pi}{2}_0 \sqrt{1-C^2\sin^2\phi}\,d\phi
\nonumber
\end{equation}
is the elliptic integral of the second kind.

\begin{figure}[htp]
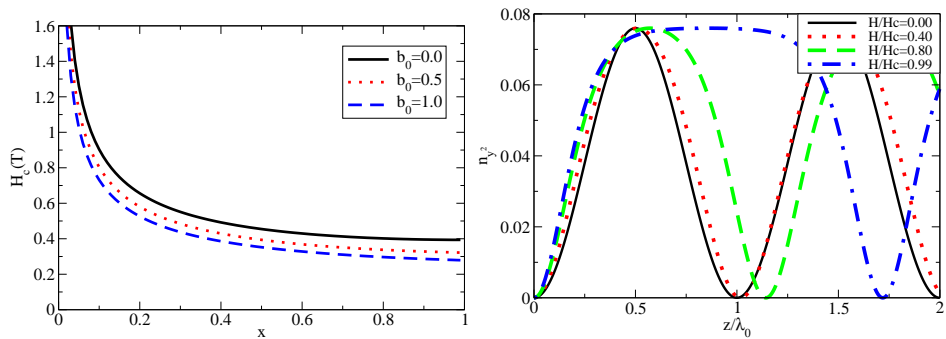

\centering
\includegraphics[scale=0.25,clip]{Hcb.eps}
\includegraphics[scale=0.25,clip]{ny.eps}
\caption{a) Critical field $H_c$ versus $x$. In the uniform nematic phase ($x=0$), there is no defined field. As $x$ varies towards the values of the cholesteric phase, the critical field monotonically decreases. Notice that as $b_0$ increases, the critical field decreases. In this figure we used $K_{22}/K_{33}=5$, $\chi_a=4 \pi\times10^{-7}$ $N/A^2$ and $q_0=3.14\times10^8$ $m^{-1}$. b) director component $n_y^2(\phi)$ versus $z/\lambda_0$ for several values of $H/H_c$. }
\label{figure:Fig3}
\end{figure}

For $C = 1$, equation~(\ref{1})  permits us to define the magnetic critical field as

\begin{equation}
\centering
H_c=\dfrac{\pi K_{22} q_0}{2\sqrt{\alpha\chi_a}}.
\label{17}
\end{equation}
Introducing $H_c$ in equation~(~\ref{1}) one finds

\begin{equation}\label{Ereduced}
\centering
\frac{H}{H_c}= \frac{C}{E_2(C)}.
\end{equation}
For $H>H_c$ the latter equation has no solution while $C$ varies from 0 to 1 when $H<H_c$.
Finally, the wavelength of the modulation under field is calculated by substituting $C$ from equation~(\ref{1}) into equation~(\ref{lambda}). Thus,

\begin{equation}\label{Lreduced}
\centering
\frac{\lambda}{\lambda_0}= \left( \frac{2}{\pi}\right)^2\,E_1(C)E_2(C),
\end{equation}
where $\lambda_0=\dfrac{\pi\alpha}{K_{22}q_0x}$ is the natural wavelength of the modulation. Equation~\ref{17}, in the case of a cholesteric $(x=1)$~\cite{pre}, reduces to the one found by de Gennes~\cite{degennes1}, when $b_0=1$. Interestingly, we notice the existence of a critical field for cholesteric phases ($x=1$) where $b_0\neq 1$. This cholesteric phase has a residual deformation from the $\rm N_{tb}$ phase, occurrence allowed by the symmetry of bent-shaped molecules. From the equations~(\ref{Ereduced}) and~(\ref{Lreduced}),  one can find 
$\lambda(H)$. In figure~\ref{figure:Fig3}-a $H_c$ as a function of $x$ is shown for some values of $b_0$ when $q_0=3.14\times10^8$ $m^{-1}$. For $x=0$ the critical field is not defined since there is no helix (uniform nematic) to unwind. As $x$ changes from 0 to 1, the critical field monotonically decreases. With the values chosen, the critical field decreases as $b_0$ increases. In figure~\ref{figure:Fig3}-b, the director component $n_y^2(\phi)$ versus $z/\lambda_0$ is shown for several values of $H/H_c$. As the field increases, the director component gets distorted, square wave like, representing the helix unwinding process. Figure~\ref{figure:Fig3}-b gives the $y$ component of the director used to plot the curves shown in figure~\ref{fig01}.

\begin{figure}[htp]
\centering
\includegraphics[scale=0.3,clip]{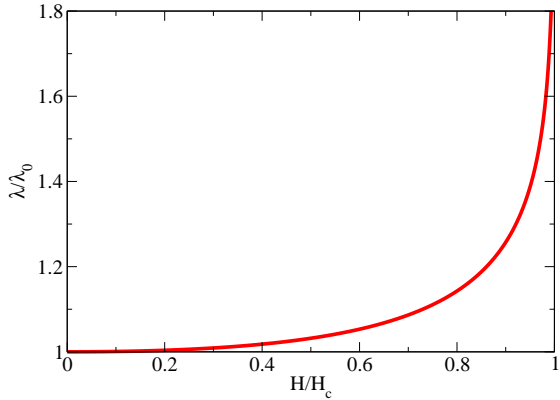}
\caption{Reduced wavelength $\lambda/\lambda_0$ of the $\rm N_{tb}$ modulation vs. the reduced magnetic field $H/H_c$. The  $\rm N_{tb}$ helix is continuously unwinded with the field,  and at $H_c$ a  transition towards a uniformly tilted nematic occurs.}
\label{figure:Fig4}
\end{figure}

Figure~\ref{figure:Fig4} shows the reduced wavelength of the modulation $\lambda/\lambda_0$ as a function of the reduced external field $H/H_c$. The calculation, in the case of an electric field $\bf{E}$,  is similar and to obtain the corresponding results one has to promote the substitution $\chi_a H^2/2 \leftrightarrow \Delta\epsilon E^2/8\pi$, where $\Delta\epsilon$ is the dielectric anisotropy. Nevertheless, in the case of an electric field one has to consider also the flexoelectric effect that is linear in the electric field, and results in a rotation of the helical axis in the plane that is normal to the the electric field direction.

\section{Conclusions}

We have theoretically investigated the effect of an applied magnetic field on the director organization of a twist-bend nematic phase. We started out by rewriting the free energy found in~\cite{pre} in terms of the parameters $q_0$ and $b_0$, associated with the natural distortions of the phase. Then, we analyzed two distinct cases: first, the magnetic field was applied parallel to the helical axis \textbf{t};  in the second case,  $\textbf{H}$ was applied perpendicularly to the direction of \textbf{t}. In both situations, $x=\sin^2(\theta)$ was the same across the sample. When $\textbf{H}\parallel \textbf{t}$, the main effect of the field, for fixed $x$, is of renormalizing the coupling parameter $\eta$, though the interval of possible values for $\eta$ remains the same. Since the coupling parameter is associated to the phase transition from $\rm N_{tb}$ to nematic, the field shifts the transition point. In another words, the presence of the field modifies the way in which the director couples with the helical axis and therefore alters the $\rm N_{tb}$ range of equilibrium. This behavior agrees with recent studies on the effect of field on the $\rm N_{tb}$ phase~\cite{Challa}. When $\textbf{H}\perp \textbf{t}$, there is a critical field corresponding to the unwinding of the helix. This critical field depends on the twist elastic constant and on the wave number as expected, but also on $x$ and on $b_0$. If $x=0$, there is no critical field, since the phase in this case is a nematic perfectly aligned. When $x=1$ and $b_0=1$, the critical field reduces to the one well known critical field of cholesterics. Interestingly, the molecular symmetry allows cholesteric phases ($x=1$) even when $b_0\neq 1$. This cholesteric phase, which has a different critical field for unwinding the helix, has a different director organization, possible with a residual spontaneous bend from the $\rm N_{tb}$ phase.

\end{document}